\documentstyle[12pt]{article}
\begin{document}
\setlength{\topmargin}{-1cm}
\setlength{\oddsidemargin}{0cm}
\setlength{\evensidemargin}{0cm}
\title{
\begin{flushright}
{\large \bf CERN-TH/96-45}
\end{flushright}
\vspace{1cm}
 {\Large\bf Symmetry Non-restoration at High Temperature and Supersymmetry}}

\author{{\bf Gia Dvali}\thanks{E-mail:
dvali@surya11.cern.ch}$~$
and {\bf K.Tamvakis}\thanks{E-mail:
tamvakis@surya11.cern.ch}\thanks{On leave of absence from
Physics Department, University of Ioannina,
Ioannina GR 45110, Greece}
\\ CERN, Theory Division, CH - 1211 Geneva 23, Switzerland\\}

\date{ }
\maketitle

\begin{abstract}
We analyse the high temperature behaviour of softly broken supersymmetric
theories taking into account the role played by effective
non-renormalizable terms generated by 
the decoupling of superheavy degrees of freedom or
the Planck scale physics. It turns out that 
discrete or continuous symmetries, spontaneously broken at
intermediate scales, may never be restored, at least up to
temperatures of the cutoff scale. There are a few interesting differences
from the usual non-restoration in non-supersymmetric theories case
where one needs at least two Higgs fields and non-restoration
takes place for a range of parameters only.
We show that with non-renormalizable interactions
taken into account the non-restoration can occur for any nonzero range
of parameters even for a single Higgs
field. We show that such theories in general solve the cosmological
domain wall problem, since the thermal production
of the dangerous domain walls is enormously suppressed.
\end{abstract}

\newpage

\subsubsection*{1. Introduction}

 The study of field theories at finite temperature\cite{kldj},\cite{w}
motivated by questions related to the
cosmological evolution of the universe,
has revealed a close analogy with many condensed matter systems.
In the considered cosmological scenarios, the broken symmetries of the
effective gauge field theory that describe particle interactions
(SM or GUT) are typically restored at high temperatures in the same way
as the rotational invariance of the ferromagnet is restored by rising its
temperature. In many cases however, as in the case of a certain
ferroelectric crystal known as Seignette salt, the gauge models
exhibit a high temperature symmetry nonrestoration\cite{w},\cite{ms}.
It turns out that for a certain range of parameter space this
effect is presented in many minimal realistic particle physics models    
with spontaneously broken  discrete and continuous symmetries
\cite{ds}-\cite{dms2}. Needless to say that such behaviour would lead to
a different picture of the hot universe and have a direct relevance
for the solution \cite{ds}\cite{dms1} of some problems
of the standard big bang
cosmology, e.g. such as the domain wall \cite{dw}
and the monopole \cite{monopole} problems.

In the modern view the effective gauge theory that
describes particle interactions below the Planck scale is a softly
broken supersymmetric theory resulting from spontaneously broken
supergravity or superstrings. It has been argued\cite{mang}
that supersymmetric theories exhibit
global and gauge symmetry restoration at high
temperatures, in contrast ordinary non-supersymmetric theories
in which both types of high temperature behaviour are possible. This is
a consequence of the more constrained nature of supersymmetric
models in which all matter interactions, Yukawa as well as scalar,
are determined by the superpotential. Note that the above is independent
of the strength of supersymmetry breaking provided that it is soft.

 The decoupling of superheavy particles as we cross their mass threshold
implies that the effective theory valid at lower energies receives
knowledge of the existence of these particles only through
non-renormalizable interactions of the light fields suppressed
by inverse powers of the superheavy mass scale. In an analogous fashion,
the effective theory resulting from supergravity or superstrings
below the Planck scale  $M_{P}$
displays an infinity of non-renormalizable
interactions suppressed by the inverse powers of the Planck mass
resulting from integrating out of the heavy modes at $M_{P}$.
In both cases the theory below the superheavy scale is described by an
effective superpotential that contains non-renormalizable interactions
of the light fields \cite{gsw},\cite{nln}.
These interactions acquire particular importance in
the case of fields with vanishing renormalizable interactions, such as
moduli fields.
The finite temperature corrections to such
a theory can be computed in the standard fashion as long as the temperature
($\Theta$) stays below the cutoff scale. The scalar potential will be
modified by the field dependent terms quadratic in $\Theta$, while
the higher powers of the temperature will be suppressed by inverse
powers of the cutoff and therefore be negligible. In the present short
letter we point out that the, inevitable, presence of the non-renormalizable
interaction in the effective field theory below $M_{P}$ can imply that
in a class of supersymmetric models the high temperature phase is
the one with a broken  symmetry. The existing proofs \cite{mang},
that the globally supersymmetric theories always possess a symmetric
high temperature ground state, are not valid when non-renormalizable
superpotentials are allowed.

 It is interesting to note that the high temperature behaviour of
such theories exhibits certain differences from the previously
studied non-supersymmetric models with high temperature symmetry
non-restoration. For example, in conventional cases the symmetry
non-restoration was observed
exclusively in a system with more than one
Higgs field and only in a certain range of the parameters.
In contrast, in the case of supersymmetric
theories with non-renormalizable terms included,
the symmetry non-restoration may occur for a single Higgs field and
for all (nonzero) values of the theory parameters (compatible with
a symmetry).

\subsubsection*{2. The Role of the Non-renormalizable Couplings}

 Consider a gauge theory with a set of the chiral superfields $\Phi^i$
in various representations of the gauge group $G$ with their matter
interactions described by a superpotential $W(\Phi)$. Supersymmetry will
be assumed to be broken by the usual soft terms in the scalar potential
\begin{equation}
m^2|\Phi^i|^2 + m^2_{ik}\Phi^i\Phi^k + c_{ikl}\Phi^i\Phi^k\Phi^l + h.c.
\end{equation}
as well as the gaugino masses ${1 \over 2}M_a\lambda^a \lambda^a + h.c.$.
The lowest order temperature corrections can be put in the form
\begin{equation}
\Delta V = {\Theta^2 \over 24} Tr[M_s^2 + M_f^+M_f + 3M_v^2]
\end{equation}
where $M_s, M_f$ and $M_v$ are scalar, fermion and gauge boson mass
matrices respectively. Since with
the above assumed soft SUSY breaking, the contribution to the
supertrace comes out to be field-independent, we may put these
corrections in the form
\begin{equation}
\Delta V = {\Theta^2 \over 16} {\rm Tr}[M_s^2 + 3M_v^2] =
{\Theta^2 \over 8} \sum \left (|{\partial^2W \over \partial\Phi^i\partial
\Phi^k}|^2 + 4g_a^2\Phi^*_i(T^aT^a)^i_k\Phi^k\right )
\end{equation}
up to field-independent terms. Note that no field-dependent SUSY
breaking term contributes. A priori there is no reason why the global
minimum of the full $\Theta \neq 0$ potential should be the symmetric one.
Restricting oneself to the renormalizable terms only, however, always leads
to a symmetric high temperature ground state\cite{mang}.

 In order to illustrate that this need not be the case when
non-renormalizable terms are included, we consider a simplest possible
example of a single superfield $\Phi$ transforming under a discrete
$Z_2$-symmetry $\Phi \rightarrow -\Phi$. In general the superpotential may
contain an infinity of even power terms compatible with the symmetry.
For simplicity we restrict ourselves to the lowest possible
non-renormalizable coupling. So the model is described by the
superpotential
\begin{equation}
W(\Phi) = -{1 \over 2} \mu \Phi^2 + {\Phi^4 \over 4!M}
\end{equation}
where M has to be understood as some large mass $\sim M_P$.
This model could come about from a renormalizable model described by
the superpotential
\begin{equation}
W(\Phi) = -{1 \over 2} \mu \Phi^2 + M_XX^2 + \lambda X\Phi^2
\end{equation}
when the field $X$ is integrated out.
Notice that by field redefinition, any complex phase of the
parameters can be simply absorbed in the overall
phase of the superpotential and thus, cannot affect 
the symmetry properties of the potential minima.
So for definiteness we assume both $\mu$ and $M$ to be  real and
positive. At $\Theta = 0$, we have a pair of non-symmetric minima
at the intermediate scale
$\Phi = \pm \sqrt{6\mu M}$ degenerate with the symmetric minimum
at the origin.
The $\Theta \neq 0$ potential (for $\Theta >> 0$) reads
\begin{equation}
V = |\Phi|^2|-\mu + {\Phi^2 \over 6M}|^2 + 
{\Theta^2 \over 8}|-\mu + {\Phi^2 \over 2M}|^2.
\end{equation}
The equation for the extremum is 
\begin{equation}
\Phi\left (-\mu + {\Phi^2 \over 2M} \right ) 
\left ({\Theta^2 \over 8M} -\mu + {\Phi^2 \over 6M} \right ) = 0
\end{equation}
Note that $\Phi$ is real in this equation. This equation has three 
solutions $\Phi = 0, \Phi^2 = 2M\mu$ and $\Phi^2 = 6M\mu$, but 
the third one exists as far as  $\Theta^2 < 8M\mu$. Thus,
at high temperature $\Theta >  \Theta_c = \sqrt{8M\mu}$,
above the intermediate scale but still below $M_P$, the only
extrema are $\Phi^2 = 0$ and $\Phi = \pm \sqrt{2M\mu}$, the second
of which was a saddle point at $\Theta = 0$. The determinants
of the curvature matrix at these points can be easily computed and are
equal to $\mu^2(\mu^2 - \Theta^4/64M^2)$ and 
$\mu^2(\Theta^4/16M^2 - 16\mu^2/9)$ respectively. Therefore, we see that
above $\Theta_c$ the symmetric minimum becomes unstable (saddle point)
and the only minimum of the theory is the one with a broken symmetry.
Note that the intermediate scale could be quite high if $\mu$ is not
very much smaller than $M$.
 
What about higher order couplings? In general, the system may include
an infinite number of non-renormalizable terms compatible with the symmetry.
In such a case the superpotential becomes
\begin{equation}
W(\Phi) = -{1 \over 2} \mu \Phi^2 + 
{\Phi^{2n} \over 2n(2n -1)M_{(n)}^{2n-3}}
\end{equation}
where the sum over $n > 1$ is assumed.  The high temperature potential now
is
\begin{equation}
V = |\Phi|^2|-\mu + {\Phi^{2n-2} \over (2n -1)M_{(n)}^{2n-3}}|^2 + 
{\Theta^2 \over 8}|-\mu + {\Phi^{2n} \over M_{(n)}^{2n-3}}|^2.
\end{equation}
To see that above a certain temperature $\Theta >> \sqrt {M\mu}$
the minimum with $\Phi \neq 0$ is the ground state, let us simply
show that at this temperature there always is a state with
$\Phi \neq 0$ which has a lower energy than the one with unbroken
discrete symmetry. For this notice that the second term has at least
one minimum (zero of the polynomial 
$-\mu + {\Phi^{2n} \over M_{(n)}^{2n-3}}$)
with $\Phi = \hat \Phi \sim \sqrt {M\mu}$. At this point the second
term vanishes (by definition) and the energy is given by the first
term which is of the order
\begin{equation}
\sim |\hat \Phi|^2\mu^2,
\end{equation}
whereas the energy of the symmetric state $\Phi = 0$ is given by the
second term and is equal to
\begin{equation}
{\Theta^2 \over 8}\mu^2
\end{equation}
Thus, above a certain temperature ($>> \sqrt{M\mu}$) the state with
a broken symmetry is lowest energy state.
 
 We can easily generalize the previous toy model into a two field $U(1)$
gauge model with a superpotential
\begin{equation}
W(\Phi,\bar{\Phi}) = -\mu (\Phi\bar{\Phi}) +
{(\Phi\bar{\Phi})^2 \over 4M}
\end{equation}
The potential now is
\begin{eqnarray}
V &=& {\Theta^2 \over 8}\left [ 2|-\mu + {\Phi\bar{\Phi} \over M}|^2
+ (|\Phi|^4 + |\bar{\Phi}|^4)/4M^2 + 
4g^2(|\Phi|^2 + |\bar{\Phi}|^2) \right ]\nonumber\\
&+& (|\Phi|^2 + |\bar{\Phi}|^2)|-\mu + {\Phi\bar{\Phi} \over 2M}|^2 +
+ {g^2 \over 2}(|\Phi|^2 - |\bar{\Phi}|^2)^2
\end{eqnarray}
Again, the $\Theta = 0$ vacua are $\Phi = \bar{\Phi} = 0$ and
$|\Phi| = |\bar{\Phi}| = \sqrt{2M\mu}$. For $\Theta \neq 0$ the potential
is minimized by $|\Phi| = |\bar{\Phi}| = v$ where either $v = 0$ or
\begin{equation}
v^4 + v^2(-8M\mu/3 + 5\theta^2/12) + 4(M\mu)^2/3 +
(-M\mu + 2g^2M^2)\Theta^2/3 = 0
\end{equation}
Clearly, the gauge term favours the symmetric minimum so
non-restoration can take
place only if $g^2 < \mu/2M$. Now at high temperatures
($>> \sqrt{M\mu}$), above the intermediate mass scale, except the symmetric
extremum at the origin we have the solution 
$|\Phi| = |\bar{\Phi}| = \sqrt{4M\mu/5}$. The curvature matrix at the
origin has eigenvalues
\begin{equation}
\mu^2( 1 \pm {\Theta^2 \over 8M\mu}) + {g^2 \over 2}\Theta^2
\end{equation}
indicating that the origin gets destabilized  for
$\Theta > \sqrt{8M\mu}$. Thus, we conclude that at high
temperature the broken phase lies lower and is the ground
state of the system.

 Let us now discuss an example with $R$-symmetry non-restoration
at high temperature. Consider the simplest
model with a discrete $R$-symmetry,
under which the superpotential changes sign $W \rightarrow - W$, and a
single superfield with the same transformation properties
$\Phi \rightarrow -\Phi$. Then, the most general superpotential is
simply a polynomial with only odd powers $\Phi^{2n + 1}$ included.
For simplicity we consider only the case with $n < 3$. Thus, the
superpotential becomes:
\begin{equation}
W = \mu^2\Phi  - {h \over 3}\Phi^3 + {\Phi^5 \over 5M^2}. 
\end{equation}
Since we are interested in zero temperature breaking of $R$ symmetry
at the scales much below $M$, we assume, as before, $\mu << M$.
Note that for $h \sim 1$ the theory still admits the zero temperature
ground state with $\Phi \sim M$. So we choose $h \sim \mu/M$ and
for convenience write it as $h = \lambda \mu/M$
where $\lambda$ is a parameter of order one. Now the
only zero temperature supersymmetric ground states are the ones
with $R$ symmetry spontaneously broken at the intermediate scale
\begin{equation}
\Phi_{\pm}^2 = {\lambda M\mu \over 2}
\left ( 1 \pm \sqrt{1 - {4 \over \lambda^2}} \right )
\end{equation}
At high $\Theta$ the potential becomes
\begin{equation}
V = {1 \over M^4}\left [ |\mu^2M^2 - \lambda \mu M \Phi^2 + \Phi^5|^2
+ 2\Theta^2|\Phi|^2|\Phi^2 - \lambda\mu M/2|^2 \right ]
\end{equation}
We see that the second, $\Theta$-dependent, term has two degenerated
minima at any temperature
\begin{equation}
\Phi = 0,~~~\Phi^2 = {\lambda \over 2}\mu M
\end{equation}
whether the minimum  with broken symmetry is a lowest one, is decided
by the first term (zero temperature potential), which splits the energies
of the above solutions, and this is simply a matter of the parameter choice.
To provide a simple existence proof let us choose $\lambda = 2$.
Then, the first term has a minimum 
$\Phi_{\pm} = \pm \sqrt{M\mu}$ which coincides with the minimum of the
$\Theta$-dependent term and thus, is a true ground state of the system!
The energy difference between this minimum and the one with an
unbroken symmetry is $\mu^4$.
So we see that in a range of parameters the $R$-symmetry is never restored.

The above examples are sufficient to make our point. The discussion of
what happens in more realistic theories, like MSSM or GUTs, is left for
a future article \cite{pr}.

\subsubsection*{3. Application: the Domain Wall Problem}

It has been shown \cite{ds},\cite{dms1}
that the symmetry non-restoration at high
temperature may provide a natural solution to the domain wall\cite{dw}
and the monopole problems \cite{monopole},
which are grave difficulties
for the standard cosmological scenario. Here we will address the issue
of the domain wall problem in the context of the models of spontaneously
broken discrete symmetries induced by the effective non-renormalizable  
couplings in the superpotential. As we have seen these systems exhibit
high temperature symmetry
non-restoration with characteristic features.
The crucial point is that the order parameter does not necessarily
grow with temperature, but may become frozen. This goes in contrast with
previously studied cases (with symmetries being broken by renormalizable
interactions) in which the order parameter to temperature
ratio remains constant at high temperature, so that the thermal production
of the domain walls is enormously suppressed\cite{ds}. This ensures that
the absence of the phase transition suffices
to solve the domain wall problem.
In our case this question needs an additional study.
For this let us consider again a simple prototype model with superpotential
(4). As was pointed out, this system
at $\Theta = 0$ has five extrema: three of them ($\Phi = 0$ and
$\Phi'_{\pm} = \pm \sqrt{6M\mu}$)
are the degenerate local minima,
and the other two $\Phi_{\pm} = \pm \sqrt{2M\mu}$ are saddle points.
At $\Theta \neq 0$ the points $\Phi'_{\pm}$ get displaced
\begin{equation}
\Phi'_{\pm} = \pm \sqrt{6\left( M\mu - {\Theta^2 \over 8}\right)}
\end{equation}
and disappear at $\Theta > \Theta_c = \sqrt{8M\mu}$. Above this critical
temperature the theory has two degenerate minima
$\Phi_{\pm} = \pm \sqrt{2M\mu}$ and one saddle point $\Phi_0 = 0$.
Thus, the symmetry is never restored above $\Theta_c$. However, for the
study of the domain wall formation, we need to consider the evolution
of the system in the interval $0 < \Theta < \Theta_c$. The evolution
goes as follows: $\Phi_0$ is a local minimum in the interval
$0 < \Theta <\Theta_c$, whereas, $\Phi_{\pm}$ becomes a minimum only
above $\Theta > \sqrt{{2 \over 3}}\Theta_c$  (for
$\Theta  < \sqrt{{2 \over 3}}\Theta_c$, the $\Phi_{\pm}$ is a saddle
point). The third pair $\Phi'_{\pm}$ is a local minimum in the interval
$\Theta < \sqrt{{2 \over 3}}\Theta_c$. Then, for
$\Theta_c > \Theta > \sqrt{{2 \over 3}}\Theta_c$ it becomes a saddle point
and, finally, disappears above $\Theta_c$. The important message is that
at $\Theta = \sqrt{{2 \over 3}}\Theta_c$ the extrema $\Phi_{\pm}$
and $\Phi'_{\pm}$ are coincident and represent an unstable turning point.
Thus for $\Theta = \sqrt{{2 \over 3}}\Theta_c$ the theory has no stable
point with broken symmetry and, therefore, in some interval 
$\Theta \sim \sqrt{{2 \over 3}}\Theta_c$ the symmetry is inevitably
restored. In the cosmological context this would lead to a restoration
of symmetry at $\Theta = 0$ (since during the cooling the system would
be trapped in the symmetric minimum). In order to have a discrete symmetry
spontaneously broken at zero temperature, we can assume
the soft SUSY breaking
negative mass term $-m^2|\Phi|^2$, possibly radiatively generated.
Then, the broken phase will be stable
for any $\Theta$ provided $|m| > \mu$. This avoids a 
troublesome phase transition, with the discrete
symmetry breaking, for all the temperatures and thus domain walls are never
formed by the Kibble mechanism \cite{kibble}.

Now, what about their thermal production?
First let us estimate when the domain walls would start to dominate
the universe assuming that there is at least one horizon size wall
at any time (temperature). At $\Theta >> \Theta_c$ the dominant
contribution to the wall energy density comes from the second
$\Theta$-dependent term in the potential of eq(6). The corresponding
wall solution (for the planar infinite wall) can be approximated by
the kink and its energy density per unit surface is
\begin{equation}
\sigma \sim {\Theta \over M}(M\mu)^{{3 \over 2}}.
\end{equation}
The thickness is
\begin{equation}
\delta \sim {M \over \Theta}{1 \over \sqrt{M\mu}}
\end{equation}
The energy of an $R$-radius wall is then given by $E_R \sim R^2\sigma
\sim R^2{\Theta \over M}(M\mu)^{{3 \over 2}}$ provided 
$R >> \delta$. Walls start to dominate when their energy density
overcomes that of the radiation. The corresponding temperature
$\Theta_d$
can be found from 
\begin{equation}
\sigma R_H^{-1} \sim \Theta^4_d,
\end{equation}
where $R_H \sim M/\Theta_d^2$ is a horizon size in radiation dominated era.
>From (22) and (24), we get $\Theta_d \sim \mu\sqrt{{\mu \over m}}$. However,
expression (22) for $\sigma$ is only valid until the temperature drops
to $\sim \Theta_c$. Below, the first term in eq(6) starts to dominate,
the wall tension becomes frozen  and the surface energy density
is $\sigma \sim \mu^2M$. So the horizon size walls (if present)
would dominate at best around  $\Theta_d \sim \mu$. 
Note that above $\Theta > \Theta_c$ the walls are wider than the horizon
$\delta > R_H$. So strictly speaking, there is no wall inside the horizon,
but rather the horizon could appear inside the wall. In the later case,
the energy density of the wall inside the horizon is simply a false
vacuum ($\Phi = 0$) energy density $\sim \theta^2\mu^2$. This makes our
upper limit on $\Theta_d$ even stronger, since now the condition that
walls dominate reads
\begin{equation}
\Theta_d^2 \mu^2 > \Theta^4  
\end{equation}
Thus, we conclude that the temperature at which infinite walls would
dominate is about $\Theta_d \sim \mu$.

 Thus, it is clear that we have to estimate the thermal production
rate of those walls which have a chance to survive and have
a horizon size at the temperature $\Theta_d$. Such problematic walls
are those that would have the size of the scales
that enter the horizon at $\Theta_d$.
Thus, the dangerous size of the walls at any $\Theta$ is the one obtained
by scaling $R_H(\Theta_d)$ back to the temperature $\Theta$.
In the radiation dominated era all the length scales evolve as a scale
factor $\sim \Theta^{-1}$ and thus, the comoving scale at temperature
$\Theta$ is given by
\begin{equation}
R_w(\Theta) \sim {M \over \Theta\Theta_d}.
\end{equation}
It is not surprising that the suppression factor for walls of that
size is enormous at any temperature. The thermal 
production rate is exponentially suppressed by the factor
$e^{-{E \over \Theta}}$.  In our case
\begin{equation}
{E \over \Theta} \sim \left ({M \over \Theta} \right)^2\sqrt{{M \over \mu}}.
\end{equation}
Even at temperatures $\sim M$ the formation rate is negligible.
Thus, dangerous walls are never produced thermally (at least below
the temperatures $\sim M_P$ where our estimates can be trusted). 

\subsubsection*{4. Summary}

In this letter we have studied a possible role of the non-renormalizable
interactions in the thermal history of supersymmetric theories. Our
results show that this role may be crucial, since the  non-renormalizable
couplings can prevent the internal symmetries from the
restoration at arbirtarily high temperature (at least up to $M_P$).
In contrast to previously observed cases, the order parameter does
not necessarily grow with temperature and can become frozen.
Also, it turns out that the symmetry nonrestoration may take place in a
case of a single Higgs superfield and for arbitarary values
of the parameters. Our observations indicate that in SUSY theories
the symmetries broken at intermediate scales by non-renormalizable
terms, in general, have a tendency to non-restoration.
 These effects are expected to have important cosmological
consequences. In particular we have shown that they may solve the
cosmological domain wall problem in SUSY theories.

\subsubsection*{Acknowledgements}

We are grateful to Goran Senjanovi\'c and Alejandra Melfo for many
usefull discussions on the high temperature symmetry nonrestoration.

\end{document}